\documentclass[pre,twocolumn,english,longbibliography,superscriptaddress]{revtex4}
\usepackage{ifthen}
\usepackage{color}
\usepackage{graphics,graphicx,epsfig}
\usepackage{epsf,epstopdf,wrapfig}
\usepackage{amssymb,amsfonts,amsmath}
\usepackage[export]{adjustbox}
\include{epsf}
\usepackage{ifthen}

\newcommand{\beqn}{\begin{eqnarray}}
\newcommand{\eeqn}{\end{eqnarray}}
\newcommand{\beq}{\begin{equation}}
\newcommand{\eeq}{\end{equation}}

\newcommand{\<}{\langle}
\renewcommand{\>}{\rangle}

\definecolor{junglegreen}{rgb}{0.16, 0.67, 0.53}
\definecolor{myrtle}{rgb}{0.13, 0.26, 0.12}
\definecolor{lincolngreen}{rgb}{0.11, 0.35, 0.02}
\definecolor{forestgreen}{rgb}{0.13, 0.55, 0.13}

\newcommand{\firstequal}{These authors contributed equally.}

\newcommand{\ch}{}

\newcommand{\figureone}{
\begin{figure}
\begin{center}
\includegraphics[width=\linewidth]{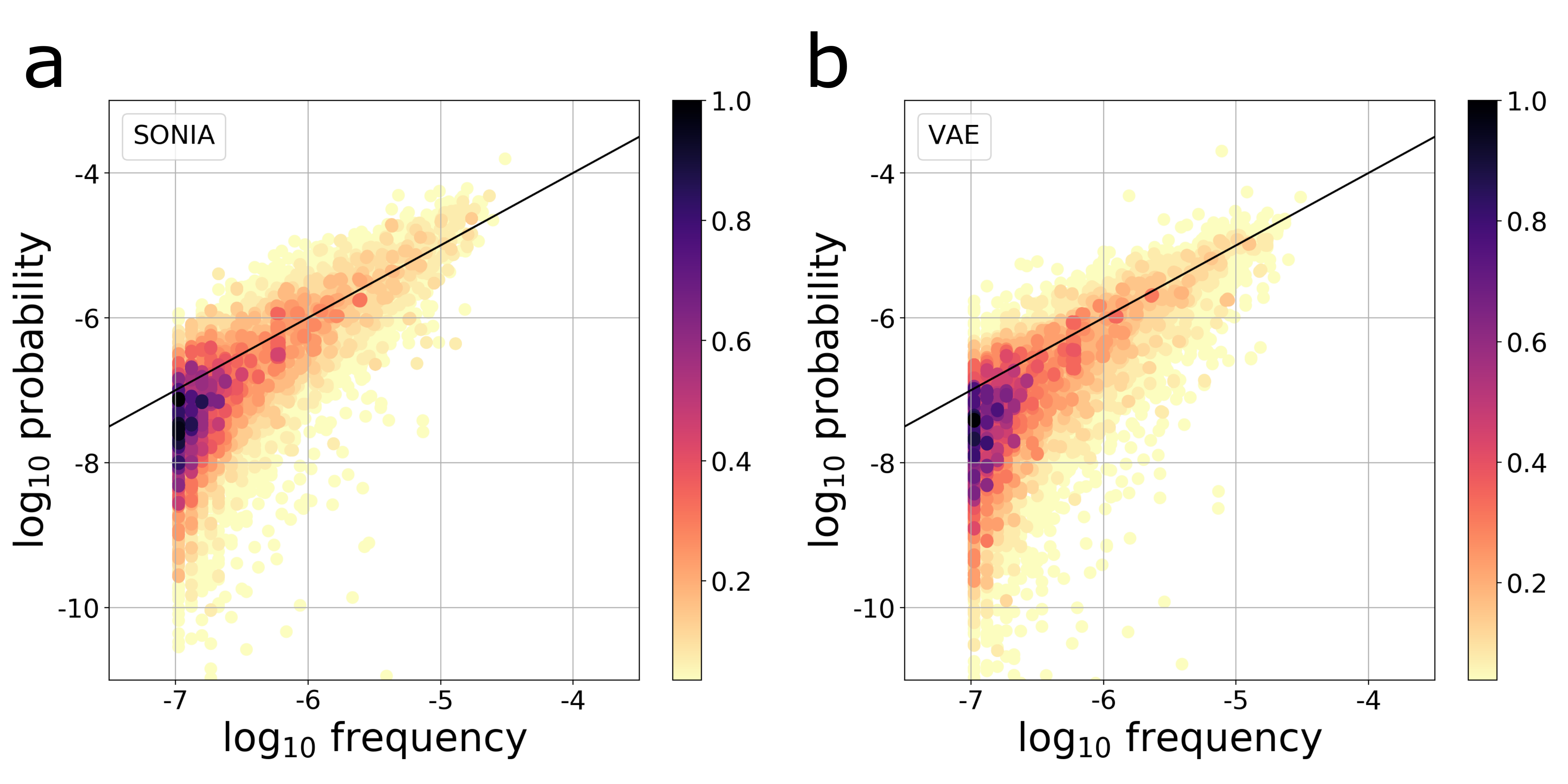}
\caption{
  Predicted TCR sequence probabilities (y-axis) versus empirical frequencies (y-axis), for (a) the SONIA Left+Right model ($\rho^2=0.53$) and (b) the VAE model ($\rho^2=0.47$). Models were trained on $2\cdot 10^5$ sequences sampled from the training set assembled from the TCR $\beta$ repertoires of 666 donors \cite{Emerson2017a}. Frequencies refer to empirical frequencies in the same datasets.
  The SONIA model was built on top of a $P_{\rm gen}$ model trained on $2\cdot 10^5$ non-productive sequences from the same donors.
}
\label{fig1}
\end{center}
\end{figure}
}

\newcommand{\figuretwo}{
\begin{figure}
\begin{center}
\includegraphics[width=\linewidth]{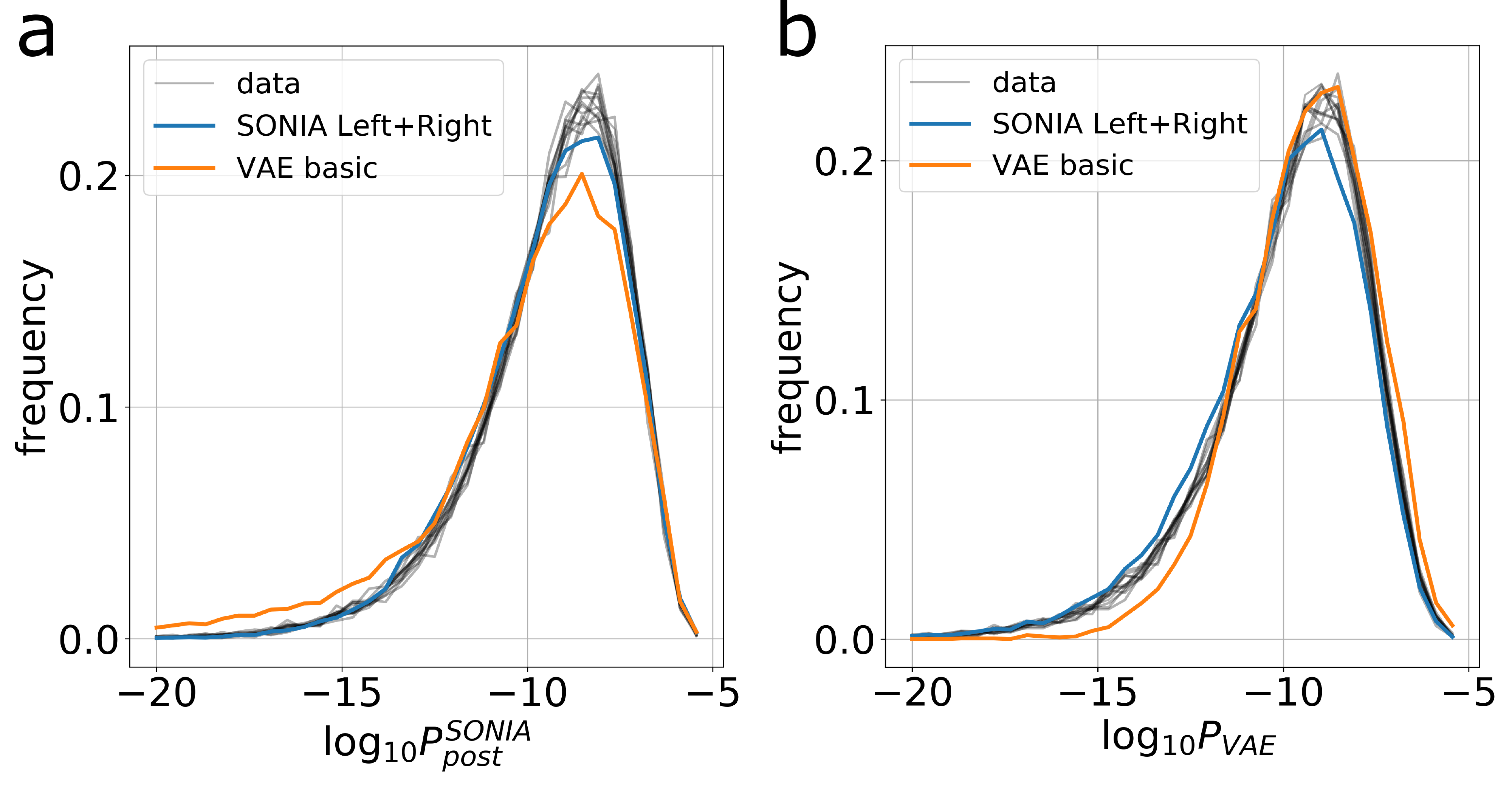}
\caption{
  (a)
  Distribution of $P_{\rm SONIA}$ of TCRs from 11 individuals from Ref.~\cite{DeNeuter2019}, as well as sequences generated by the SONIA Left+Right model, and the VAE.
  The SONIA model was trained on a set of $10^5$ sequences, on top of the $P_{\rm gen}$ model trained for Fig.~\ref{fig1}.
  (b) Distribution of $P_{\rm VAE}$ for the same sequences as in (a).
}
\label{fig2}
\end{center}
\end{figure}
}

\begin{document}
\title{On generative models of T-cell receptor sequences}

\author{Giulio Isacchini}
\thanks{\firstequal}
\affiliation{Max Planck Institute for Dynamics and Self-organization, Am Fa\ss berg 17, 37077 G\"ottingen, Germany}
\affiliation{Laboratoire de physique de l'\'Ecole normale sup\'erieure
  (PSL University), CNRS, Sorbonne Universit\'e, and Universit\'e de
  Paris, 75005 Paris, France}
\author{Zachary Sethna}
\thanks{\firstequal}
\affiliation{Memorial Sloan Kettering Cancer Center, New York, New York 10065, USA}
\author{Yuval Elhanati}
\affiliation{Memorial Sloan Kettering Cancer Center, New York, New York 10065, USA}
\author{Armita Nourmohammad}
\affiliation{Max Planck Institute for Dynamics and Self-organization, Am Fa\ss berg 17, 37077 G\"ottingen, Germany}
\affiliation{Department of Physics, University of Washington, 3910 15th Avenue Northeast, Seattle, WA 98195, USA}
\affiliation{Fred Hutchinson cancer Research Center, 1100 Fairview ave N, Seattle, WA 98109, USA}
\author{Aleksandra M. Walczak}
\affiliation{Laboratoire de physique de l'\'Ecole normale sup\'erieure (PSL University), CNRS, Sorbonne Universit\'e, and Universit\'e de Paris, 75005 Paris, France}
\author{Thierry Mora}
\affiliation{Laboratoire de physique de l'\'Ecole normale sup\'erieure
  (PSL University), CNRS, Sorbonne Universit\'e, and Universit\'e de
  Paris, 75005 Paris, France}

\begin{abstract}
T-cell receptors (TCR) are key proteins of the adaptive immune system, generated randomly in each individual, whose diversity underlies our ability to recognize infections and malignancies. Modeling the distribution of TCR sequences is of key importance for immunology and medical applications. Here, we compare two inference methods trained on high-throughput sequencing data: a knowledge-guided approach, which accounts for the details of sequence generation, supplemented by a physics-inspired model of selection; and a knowledge-free Variational Auto-Encoder based on deep artificial neural networks. We show that the knowledge-guided model outperforms the deep network approach at predicting TCR probabilities, while being more interpretable, at a lower computational cost.
\end{abstract}

\maketitle

\section{Introduction}

Deep learning methods are proving a very useful approach in many areas of physics and the natural sciences \cite{Ching2018,Carleo2019}. These algorithms are successful in identifying hidden patterns in large amounts of data, often helping make progress in situations where traditional analyses reach their limits~\cite{Yamins2016,Senior2020,Yang2019}. Despite the black box aspect of how the algorithm works and the lack of interpretability of the model features, machine learning is undoubtedly useful, especially in cases where the natural system of interest escapes our intuition or knowledge. However, as we show here on the example of immune repertoires, introducing physical or biological intuition into data-driven models can outperform basic uninformed machined learning approaches.

The adaptive immune system is made up of a large ensemble of diverse lymphocyte receptors that recognize different pathogens. The receptors expressed on the surface of T cells (T cell receptors - TCR) are generated by randomly assembling genomic templates for three genes (variable---V, diverse---D and junction---J) that make up of the so-called $\beta$ chain, and two (V and J) for the $\alpha$ chain. Additionally to this combinatoric diversity, non-templated nucleotides are added at the junctions between these templates and nucleotides are deleted. Such recombined DNA forms the newly generated TCR that later undergoes thymic selection that tests for its ability to form a receptor protein and bind, albeit not too strongly, proteins that are natural to the host organism \cite{Yates2014}. TCR that pass thymic selection are released into the periphery and form the naive repertoire (i.e. non-stimulated by foreign antigens). Due to the random addition and deletion of nucleotides, receptors sequences have different  lengths and some are even out of frame or have stop codons, in which case they are called nonproductive. Conversely, sequences with no frameshift nor stop codon are conventionally called productive. High-throughput immune repertoire sequencing experiments sample blood from individual hosts, sort out TCRs, and sequence this subset~\cite{Heather2017,Minervina2019a,Bradley2019}. Analysis of this kind of data makes it possible to characterize the statistics of both generated and naive repertoires.

TCR sequences differ from classical protein families, which are grouped by function and across species \cite{El-gebali2019}. Those families are believed to have evolved over long time scales under a shared selective pressure that shapes their statistics. For such families, physics-inspired statistical inference methods have helped to predict contacts between amino-acids in the protein \cite{Cocco2018}, define sectors of co-evolving residues \cite{Halabi2009}, or find interaction partners \cite{Bitbol2016a}. Deep \cite{Riesselman2018} and non-deep \cite{Tubiana2019} machine learning approaches have also been successfully applied. By contrast, TCR generation is fairly well understood mechanistically.
Previously we developed a statistical inference technique that uses biological knowledge of the underlying assembly processes to learn the statistics of generation and calculate the generation probability of each TCR sequence~\cite{Murugan2012,Marcou2018}. Since thymic selection involves many specific interactions with antigen-presenting cells, modeling it from first principles is more difficult. Nevertheless, simple models of selection based on the assumption of an additive fitness \cite{Berg1987a} have been shown to well recapitulate some key statistics of these ensembles \cite{Elhanati2014,Sethna2020}. However, a direct test of the performance of this method for the abundance of specific sequences in large cohorts is still lacking.

Recently, Davidsen et al. \cite{Davidsen2018} described an elegant approach for learning the distribution of T-cell receptor beta sequences (TCR$\beta$ or simply TCR in the following), based on a Variational Auto-Encoder (VAE). The method makes it possible to generate new sequences with the same statistics as real repertoires, and to evaluate the frequency of individual sequences, which agree with the data with good accuracy. Its main strength is that it does not take any information about the origin of these sequences through VDJ recombination and thymic and peripheral selection. Yet it manages to extract statistical regularities imprinted by these processes.

Here we compare the VAE method \cite{Davidsen2018} with the previously proposed model of generation and selection, called SONIA \cite{Elhanati2014,Sethna2020}. We compare their performances for predicting the distribution of TCR sequences in controlled conditions, training and validating on the same datasets. Contrary to the claims of the original VAE paper \cite{Davidsen2018}, we show that that knowledge guided models perform as well as the variational auto-encoder or even better, at a lower computational cost.

\section{Model definitions}

\subsection{Knowledge-guided model}

To predict the probability distribution of TCR sequences, we build a generative model that proceeds in two steps: initial generation, and selection.

First, a recombination model for the probability of generation of a sequence $\sigma$, denoted by $P_{\rm gen}(\sigma)$, is learned from failed, nonproductive rearrangements, which are free of selection biases \cite{Murugan2012,Marcou2018}. This model describes in detail the probabilities of V, D, and J usages, and of deletion and insertion profiles. Calling $E$ the collective variable describing the recombination scenario, the model predicts its probability $P_{\rm scenario}(E)$. Its parameters are learned through Expectation-Maximization using the IGoR software \cite{Marcou2018}.

Although the model is trained on non-productive sequences, it can be used to predict the probability of any sequence.
Denoting $\hat\sigma(E)$ the amino-acid sequence produced by scenario $E$, we define the generation probability of a productive amino-acid sequence $\sigma$ as:
\beq\label{pgen}
P_{\rm gen}(\sigma)=\frac{1}{F}\sum_{E}P_{\rm scenario}(E) \mathbb{I}[\hat\sigma(E)=\sigma],
\eeq
where $\mathbb{I}(\cdot)$ is the indicator function, and $F=\sum_EP_{\rm scenario}(E)\mathbb{I}(\hat\sigma(E)\textrm{ is productive})$ is the probability that a random recombination scenario results in a productive sequence. More precisely, $\sigma$ is defined by the choice of $V$ and $J$ genes ($\sigma_V$ and $\sigma_J$), as well as the amino-acid sequence of the Complementarity Determining Region 3 (CDR3) that lies between $V$ and $J$, $\sigma_1,\ldots,\sigma_L$. The sum in \eqref{pgen} involves a large number of terms due to the degeneracy of both the genetic code and the recombination process, but it can be done using a recursive technique akin to transfer matrices implemented in the OLGA software \cite{Sethna2019}.

Second, a model of selection, called SONIA \cite{Sethna2020}, is learned on top of the generation probability $P_{\rm gen}$ to describe the distribution of productive sequences,
\beq
P_{\rm SONIA}(\sigma)=Q(\sigma)P_{\rm gen}(\sigma),
\eeq
where
\beq
Q(\sigma)=\frac{1}{Z}\exp\left[h_{VJL}(\sigma_V,\sigma_J,L)+\sum_{i=1}^L h_{i,L}(\sigma_i)\right]
\eeq
is a selection factor calculated through additive ``fields'' $h$ acting on the sequence elements, similarly to additive position-weight matrix models first introduced for DNA binding sites \cite{Berg1987a}.

Within this framework, we can define three models according to the parametrization of $h$. In the first two models, the VJL field is decomposed as $h_{VJL}(\sigma_V,\sigma_J,L)=h_{VJ}(\sigma_V,\sigma_J)+h_{L}(L)$. A first model in which $h_{i,L}$ is left unconstrained is called the ``Length-Position'' (LP) model. This choice corresponds to the original model of \cite{Elhanati2014}, in which the selective pressure on each amino-acid may depend on the sequence length $L$. However, observations \cite{Elhanati2014} suggest that these factors are to some extend independent of $L$.
This invariance can be incorporated by assuming that the field can be decomposed into two contributions depending on the position of the amino acid from the right and left ends of the CDR3: $h_{i,L}=h_{i,\rm right}+h_{L-i+1,\rm left}$. The resulting ``Left+Right'' (LR) model has much fewer parameters and is less likely to overfit the data. For these two models, parameters are learned by maximizing the log-likelihood with an $L^2$ regularization using gradient ascent, as specified in Ref.~\cite{Sethna2020}.

In addition, because no
software implementation of the selection model was provided with the original article \cite{Elhanati2014}, Davidsen et al. \cite{Davidsen2018} compared their VAE approach to a {\ch reduced} version of this selection model (not examined in
\cite{Elhanati2014}), which they call \texttt{OLGA.Q}. In that model, only VJ usage and CDR3 length were included: $h_{i,L}=0$. Its parameters $h_{VJL}$ were fitted by maximizing the likelihood analytically.

\subsection{Variational Auto-Encoder}

A VAE is an auto-encoder whose structure can be used as a generative probabilistic model. A good introduction can be found in Ref.~\cite{Kingma2019}. In short, a VAE consists of a probabilistic encoder $q(z|\sigma)$ and a probabilistic decoder $p(\sigma|z)$, converting the sequence into a continuous multi-dimensional latent variable $z$ and back. The goal of the encoder is to make the probabilistic mapping from $\sigma$ to itself through $q$ and $p$ as faithful as possible, while at the same time making the distribution of the latent variable $z$ as close as possible to a simple distribution, i.e. multivariate Gaussian with unit covariance.

Both $p$ and $q$ are parametrized by deep neural networks, whose parameters are optimized for these two objectives, using stochastic gradient descent. Once the model is learned, new sequences can be generated by drawing $z$ from $p_0(z)$, and $\sigma$ from $p(\sigma|z)$, so that $\sigma$ is distributed according to $P_{\rm VAE}(\sigma)=\int dz\, p(\sigma|z)p_0(z)$. In practice, the predicted probability of a given sequence $P_{\rm VAE}(\sigma)$ is evaluated using Monte-Carlo importance sampling.
In Ref.~\cite{Davidsen2018}, a variant of the traditional auto-encoder detailed in \cite{Higgins2017} was used. Here we focus on the version of the VAE called \texttt{basic} in that paper.

\figureone

\figuretwo

\begin{table}
\centering
\begin{tabular}{c|c|c|c|c}
 &$10^6$ $\rho^2$& $10^6$ $D_{KL}$ & $2\cdot 10^5$ $\rho^2$ & $2\cdot 10^5$ $D_{KL}$\\ 
\hline
VAE & 0.48 & 1.7 & 0.47 & 2.0 \\ 
$P_{\rm gen}$  & 0.48 & 4.5 & 0.51 & 4.5\\  
\texttt{OLGA.Q} & 0.48 & 2.6 & 0.47 & 2.6\\
SONIA LP & 0.52 & 1.8 &  0.52 & 1.7 \\
SONIA LR & 0.53 & 1.4 &  0.53 & 1.4\\
\end{tabular}
\caption{Pearson's correlation coefficients $\rho^2$ and Kullback-Leibler divergence $D_{KL}$ (in bits) for the various models. Either 1 million or 200,000 sequences were used in the training dataset.}
\label{table}
\end{table}

\section{Model comparison}

\subsection{Datasets and model training}

The data consists of TCR$\beta$ sequence repertoires of 666 individuals \cite{Emerson2017a}. We use the exact same procedure, dataset, and subsamples as in \cite{Davidsen2018} for reproducibility.

For each individual, read counts are first discarded as they stem from clonal expansions.
To train an initial $P_{\rm gen}$ model on which SONIA is built and trained, we used $2\cdot 10^5$ nonproductive sequences drawn randomly from all donors.
For all models, unique amino-acid sequences were first separated into a training and a testing dataset of equal sizes.
All models were then trained on $2\cdot 10^5$ or $10^6$ TCR$\beta$ sequences randomly sampled from the training dataset with replacement, according to their frequency in the cohort, counting each unique nucleotide sequence in each patient. Their performance was assessed by their ability to predict the frequency of sequences from the testing set, $P_{\rm data}(\sigma)$.

\subsection{Predicting sequence frequencies}

We used two measures of performance: Pearson's $\rho^2$ between the logarithms of the frequencies as in \cite{Davidsen2018}, and the Kullback-Leibler divergence: $D_{KL}=\<\log_2[P_{\rm data}(\sigma)/P_{\rm model}(\sigma)]\>$ (${\rm model}={\rm VAE}$ or SONIA), where the average $\<\cdot\>$ is taken over $10^4$ sequences from the testing set, sampled according to their relative frequencies within that set. We excluded $\sim 0.3\%$ of sequences for which $P_{\rm gen}=0$, probably due to sequencing errors. Note that, if not for the $L^2$ regularization, maximizing the log-likelihood would be equivalent to minimizing the $D_{KL}$. The scale of $D_{\rm KL}$ may be compared to the total entropy of the ensemble, $-\sum_\sigma P_{\rm SONIA}(\sigma)\log_2 P_{\rm SONIA}(\sigma)\approx 31$ bits \cite{Sethna2020}.

Fig.~\ref{fig1} shows the predicted frequencies of the Left+Right SONIA model and the VAE model, both trained on the same $2\cdot 10^5$ sequences, and compares them to data. The performances of all models and both datasets are reported in Table \ref{table}. SONIA models perform generally better than the VAE, especially the Left+Right model which is the best model according to both measures of performance. Note that the Length-Position model of Ref.~\cite{Elhanati2014}, also performs as well as the VAE.
Davidsen et al. \cite{Davidsen2018} did not compare their model to it owing to the absence of a readily available implementation.

Strikingly, even the basic model of generation with no selection ($h=0$), $P_{\rm gen}$, performs comparably to the VAE, and sometimes better according to the $\rho^2$ measure, despite the model being trained on nonproductive sequences. Accordingly, the \texttt{OLGA.Q} model, which adds a minimal layer of selection on top of $P_{\rm gen}$, also performs very well. These results differ substantially from the $\rho^2=0.26$ - $0.27$ reported in \cite{Davidsen2018} for \texttt{OLGA.Q}. In \cite{Davidsen2018}, the default model for $P_{\rm gen}$ was not actually trained on the dataset of interest, but rather used with its default parameters learned from a different dataset, which explains the poor reported performance.

We can also compare the two models by asking whether the distribution of frequencies are well reproduced by one another, using another TCR dataset from \cite{DeNeuter2019} to allow for a direct comparison to the results of Ref.~\cite{Davidsen2018} (Fig.~\ref{fig2}).
Both the VAE and SONIA agree with the data in their distribution of $P_{\rm model}$.
VAE-generated sequences have the same distribution of $P_{\rm SONIA}$ as SONIA-generated sequences, with a slight under-estimation of the distribution peak, and an excess of low-frequency sequences (Fig.~\ref{fig2}a). The converse is true when looking at the distribution of $P_{\rm VAE}$ for SONIA- versus VAE-generated sequences (Fig.~\ref{fig2}b). This suggests that the VAE and SONIA capture some features of the sequence statistics that are distinct from one another.

\subsection{Computational times}
SONIA is an order of magnitude faster than the VAE, which uses Monte-Carlo sampling to calculate predicted frequencies. The average computing time for $P_{\rm SONIA}(\sigma)$ is 14\,ms per sequence on a laptop computer and 3\,ms on a 16-core computer, versus 0.18\,s for $P_{\rm VAE}(\sigma)$ (no parallelization possible).

SONIA was also faster to train. It took 33 minutes to train a SONIA model on $10^6$ sequences using a 30-core computer, to which one should add 31 minutes to train an IGoR model on $2\cdot 10^5$ nonproductive sequences. For the same amount of data and on the same machine, the VAE took 7 hours to train.

\section{Conclusion}

In summary, both approaches, VAE and SONIA, perform equally well, with
perhaps a slight advantage for the latter. SONIA is also much faster. These results suggest that, while knowledge-free approaches such as the VAE perform well, there is still value in preserving the structure implied by the VDJ recombination process as a baseline for learning complex distributions of immune repertoires. Extending the SONIA model considered here beyond a simple linear combination of features, and taking ideas from the modeling strategy of the VAE, offers interesting directions for future improvement in repertoire modeling.

In a more general context, while machine learning approaches are undoubtably a very useful tool, they can be made even more powerful when combined with models that describe the underlying physics or biology. This is the case when training data is limited, as has been reported in complex image processing of non-animate matter \cite{Colas2019}. As we show, even if data is abundant, using models to guide learning can help. 

{\bf Code availability.} All code for reproducing the figures of this
comment can be found at
\url{https://github.com/statbiophys/compare_selection_models_2019/}. The
SONIA package upon which that code builds is available at \url{https://github.com/statbiophys/SONIA/} \cite{Sethna2020}.

{\bf Acknowledgements.} The work of GI, TM and AMW was supported
by grant European Research Council COG 724208.

\bibliographystyle{pnas}

\end{document}